\documentstyle[twoside,fleqn,espcrc2,epsfig]{article}

\newcommand{\AmS}{{\protect\the\textfont2
  A\kern-.1667em\lower.5ex\hbox{M}\kern-.125emS}}

\title{Observations of Circinus X-1 with RXTE}

\author{Hale Bradt
	\address{Center for Space Research and Department of Physics\\
	Massachusetts Institute of Technology \\ Room 37-587\\
	Cambridge MA 02178 USA},
Robert Shirey \& Alan Levine }
       
\begin{document}

\begin{abstract}
Data accumulated with RXTE during the active state of Cir~X-1 as well
as during an unusually long transition from the active state to the
quiescent state are reported. The long decline from the active state
allowed the source characteristics to be studied systematically as a
function of intensity. The following results are presented:
(1)~spectral fits during entry into a dip clearly show absorption with
partial covering as previously reported (Brandt et~al.\ 1996), and (2)
correlations between position in the hardness-intensity plane and the
character of the power density spectrum as the source entered the
quiescent state are suggestive of Z source behavior seen in LMXB
sources.
 
\end{abstract}

\maketitle

\section{INTRODUCTION}

Circinus X-1 is a highly variable X-ray source that exhibits periodic
activity every 16.6 d which is widely attributed to an eccentric
binary orbit with the activity occurring at periastron in X-rays,
infrared/optical, and radio (Kaluzienski et~al. 1976; Glass 1978;
Whelan et~al. 1977; Moneti 1992). The detection of bursts with EXOSAT,
almost surely of type I, indicates the source is very likely a
low-mass binary (Tennant, Fabian \& Shafer 1986); see also Glass
(1994).  It has also been suggested that the binary was ejected from a
nearby supernova (which formed the neutron star) and is thereby very
young (30~000 to 100~000~y; Stewart et~al. 1993).  A young age would
be in accord with the eccentric orbit. Also, quasi-periodic
oscillations (QPO) have been reported in EXOSAT data by Tennant et~al
(1987, 1988) and others (see Oosterbroek et~al. 1995) at 1.4~Hz,
5--20~Hz and 100--200~Hz.

More recently, a study with ASCA of a transition from a low state of
Cir~X-1 to a high state showed it to be caused by absorption with
partial covering (Brandt et~al. 1996). RXTE observations of the
quiescent phase of one 16-d cycle showed strong correlations in RXTE
data between various features in the PDS including a quasi periodic
oscillation that moved from 1.3--12~Hz during the quiescent part of
the cycle (Shirey et~al. 1996).  A reanalysis of EXOSAT observations
led Oosterbroek et~al. to propose that the neutron star has a very low
magnetic field with, on occasion, an accretion rate that reaches the
Eddington level; see also van der Klis (1994).

Here we present new RXTE results on the nature of the 16.6-d cycles,
the absorption in dips, and the QPO in the quiescent and active
phases.  The work presented briefly here will be presented more fully
in a forthcoming publication by Shirey and colleagues.

\begin{figure*}
\epsfig{figure=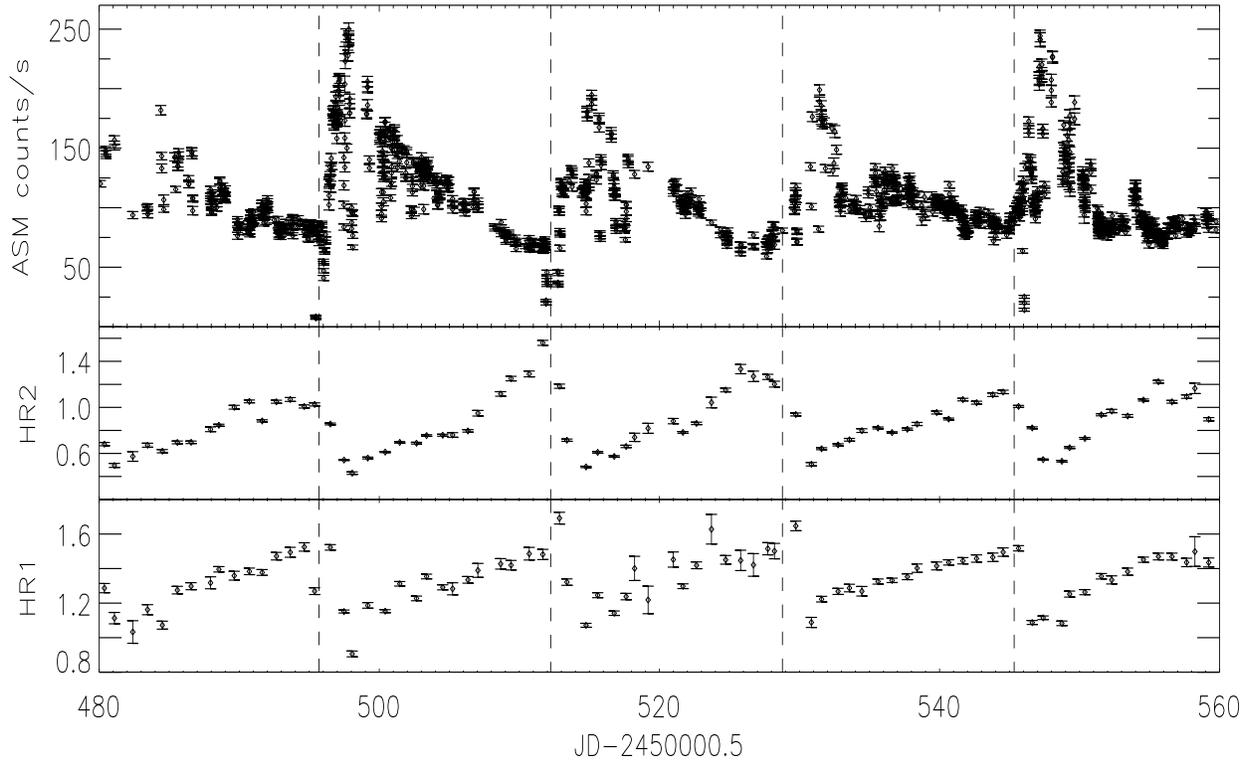,height=16.0cm,width=16.0cm,angle=+90}
\caption{
ASM light curve (1.5--12~keV) and hardness ratios
for 80 days. HR1 is 3-5/1.5-3 and HR2 is 5-12/3-5, all in keV.
Dashed lines are zero-phase markers (Stewart et~al. 1991). 
Eight PCA observations were carried out 
during the 16-d cycle (1997 Feb. 16 - Mar. 6, or at Day~$\sim$500) which 
exhibits the slow descent.
}
\label{fig:1}
\end{figure*}

\section{RESULTS}

\subsection{Long-term behavior - ASM}

The record of Cir~X-1 from the All-Sky Monitor (1.5--12~keV) of RXTE
shows it to have been remarkably consistent in its behavior over the
past 20 months. A sample ASM light curve covering five of the 16.5-d
cycles is shown in Fig.~1. The quiescent phases have remained at about
1.0~Crab. This is unusual in the context of past epochs when Cir~X-1
has been below thresholds of detection or weaker and less active. In
other words, it exhibits significant long-term (years) variability
(e.g. Oosterbroek et~al.).  In the ASM data, the active phases show
increased fluxes, a generally softer spectrum, and chaotic behavior
(Fig.~1). These features were generally known before but not in such a
comprehensive context. The figure also shows the general hardening of
the spectrum during the quiescent phases previously reported (Shirey
et~al.) and earlier observed in folded data with the Ginga ASM data
(Tsunemi et~al. 1989).

The duration and level of activity during the active states can vary
significantly from one cycle to another (Fig.~1). In particular we
note the active state occurring at day~500 in Fig.~1. It was unusually
slow in its descent to the quiescent phase. Such a slow descent allows
one to study the spectral and temporal behavior as a function of
luminosity and/or state.

\subsection{High-statistics PCA Observations}

A major study of Cir~X-1 with the RXTE PCA has been undertaken by the
present authors. The source has been studied during several active
phases with high (about 60 percent) coverage and for several complete
16-d cycles, typically with sampling every two days. To date the
observing time amounts to 840 ks. These data, soon becoming public,
exhibit Cir~X-1 undergoing a wide variety of spectral/temporal states.

PCA observations taken on 1996 Sept. 21 exhibited a dramatic dip which
is presented herein. (There are others in the archive.) Also, eight
observations at 2-day intervals were obtained (fortuitously) during
the 1997 Feb-Mar cycle wherein there was a slow descent from the
active state as noted above. The results from these latter
observations are presented here also.  Each of these eight
observations lasted for ~6 ks; they are designated Obs. I - VIII in
time order.
 
\begin{figure*}
\epsfig{figure=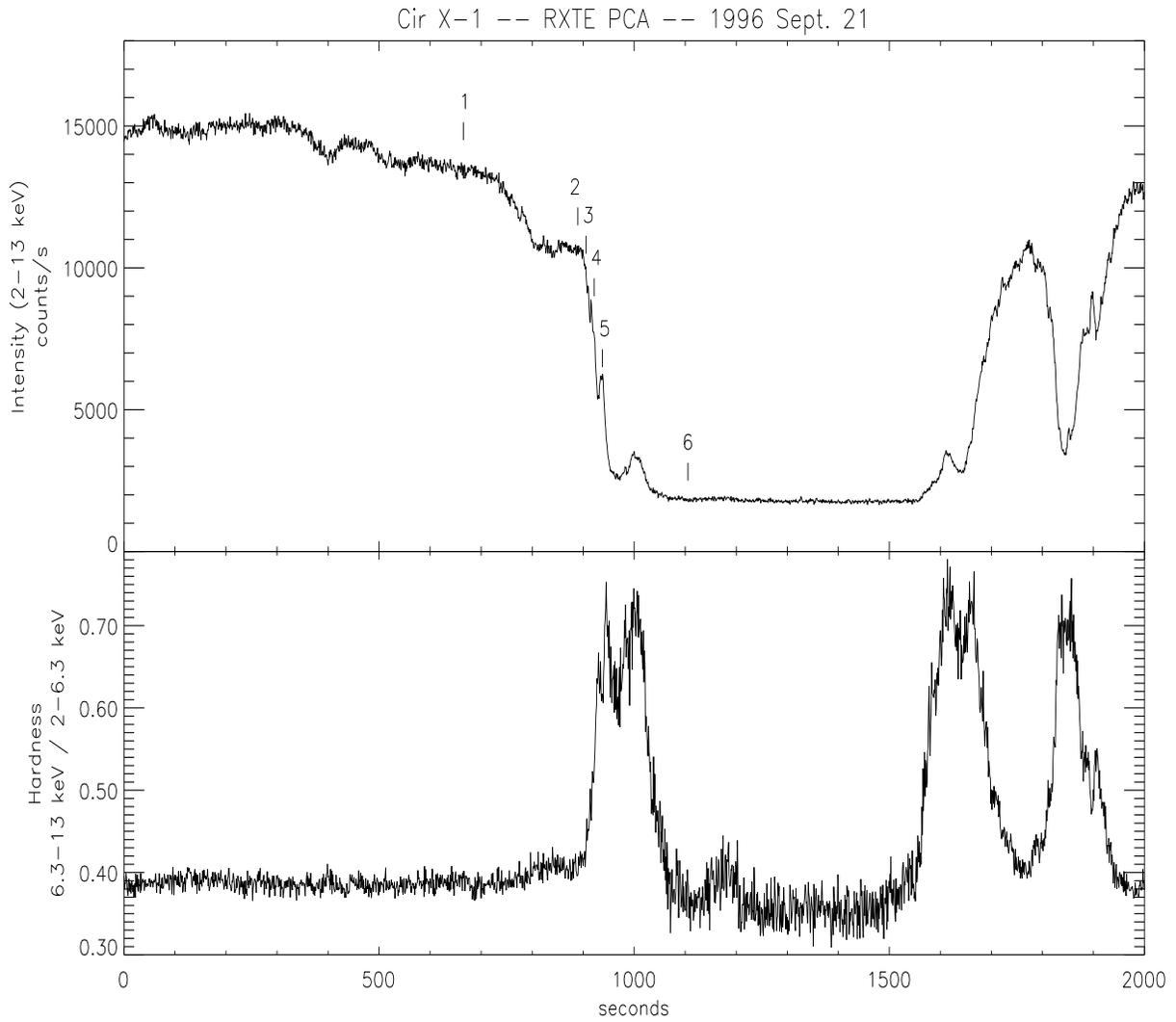,height=15.0cm,width=16.0cm,angle=+90}
\caption{
Dip of duration $\sim$800~s in PCA data, from 1996 Sept. 21.}
\label{fig:2}
\end{figure*}

\begin{figure*}
\epsfig{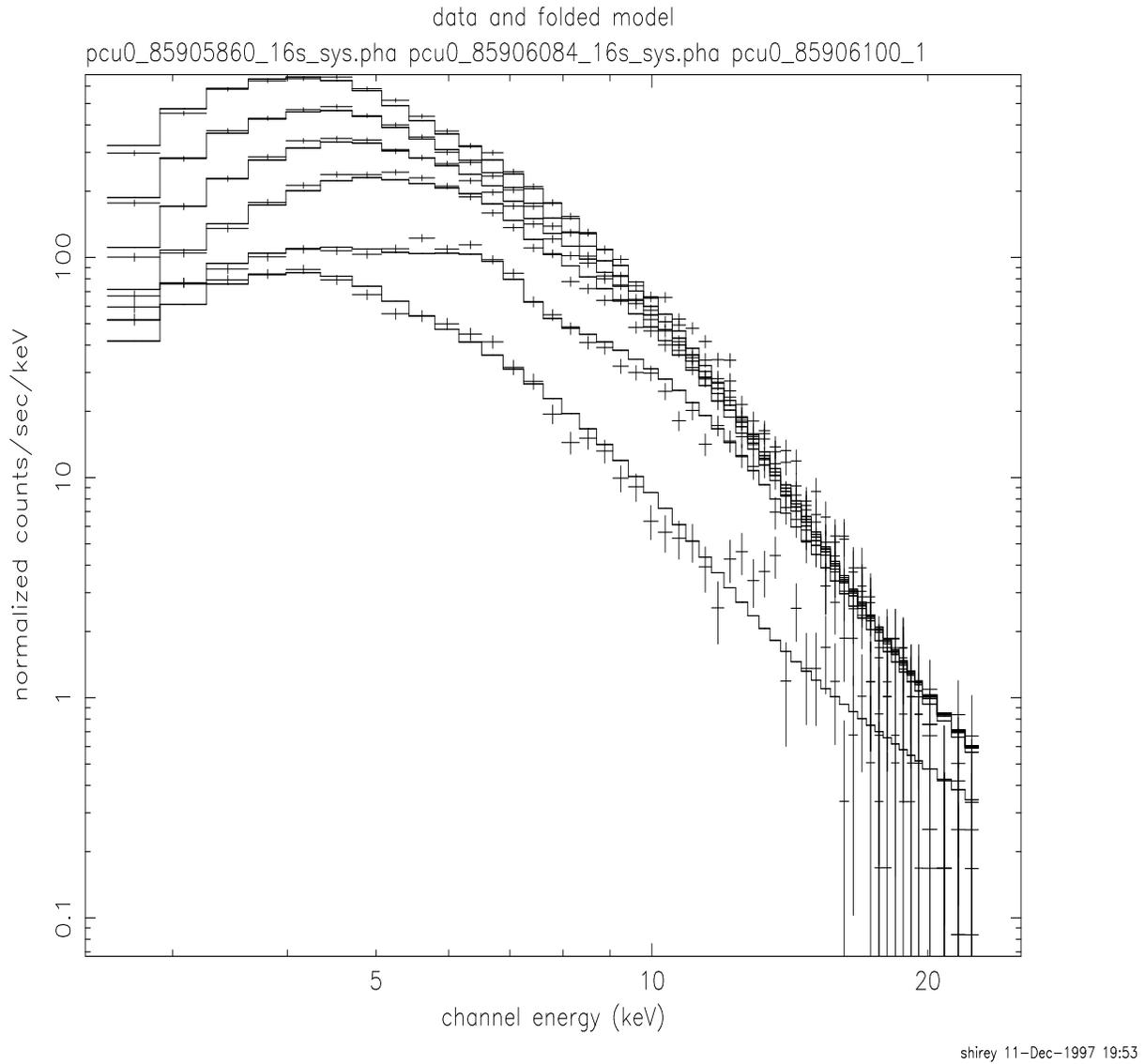}
\caption{
Data and spectral fits for the times indicated by the
numerals 1 to 6 in Fig.~2. Spectrum 1 is the uppermost and spectra 2 -
6 sequentially decrease at 4 - 6 keV.}
\label{fig:3}
\end{figure*}

\subsection{Dips}

Low states or dips are seen in Fig.~1 to occur very close to phase
zero. They systematically precede the subsequent increase of flux and
flaring behavior of the active state. They should be an important
indicator of the geometry of the matter in the binary system.

The dips are beautifully elucidated in the PCA data with sufficient
statistics to follow the spectra during the transition into the
dip. One such dip on 1996 Sept. 21 is shown in Fig.~2 with a hardness
ratio. Note the hardening during entry and the softness during the
minimum.

The six times marked in Fig.~2 were chosen for the determination of
energy spectra, one before entry into the dip, four during the
descent, and one in the dip at low flux. They are numbered
sequentially from 1 to 6. (Do not confuse with the observation numbers
I - VIII used for the 1997 data.)  The six count-rate spectra are
shown in Fig.~3, together with fits through the data points. The
spectrum taken at time 1 is uppermost. The underlying continuum model
is a disk-blackbody plus a power law with absorption of 2 x 10$^{22}$
cm$^{-2}$ (Predahl and Schmitt 1995; Brandt et~al.). This
(interstellar) absorption was held fixed for the fits. The PL and DBB
components were allowed to vary but were constrained to be the same
for all six spectra. In addition, the model included partial
covering. The parameters for this were allowed to vary, namely the
column density and fraction covered.

The fits give do not give satisfactory chi square values because of
the imposed constraints, but they do provide qualitative insight into
the transition. The partial covering is evident in that the flux in
the lowest channel reaches a minimum in spectrum 4 and is
approximately constant thereafter.  The absorption has become so great
that, in the covered region, the low-energy flux is completely
absorbed, and only the uncovered flux remains. Thereafter the
absorption affects only higher and higher energies, with iron
absorption being pronounced as one can see from the structure in the
lower two spectra, \#5 and \#6.

The covering fraction was set to zero for spectrum \#1. The fits for
spectra \#2 to \#6 yield values increasing monotonically from 0.45 to
0.87. The major differences in the spectra are the column densities of
the partial covering. They too show a monotonic increase from zero for
spectrum \#1 and then from 17 +/- 1.5 up to 675 +/- 102 for
observations \#2 to \#6, in units of 10$^{22}$ cm$^{-2}$. Note that
spectrum \#6 has N$_{H}$ exceeding 10$^{24}$ cm$^{-2}$. This is in
accord with the absorption found by Brandt et~al.

Although this absorption was reported earlier, it is reassuring to
find that this broad-band spectral approach provides similar
conclusions. Also, this work places the absorption clearly in the
context of the well-defined dips seen in the ASM and PCA data.

\begin{figure*}
\epsfig{figure=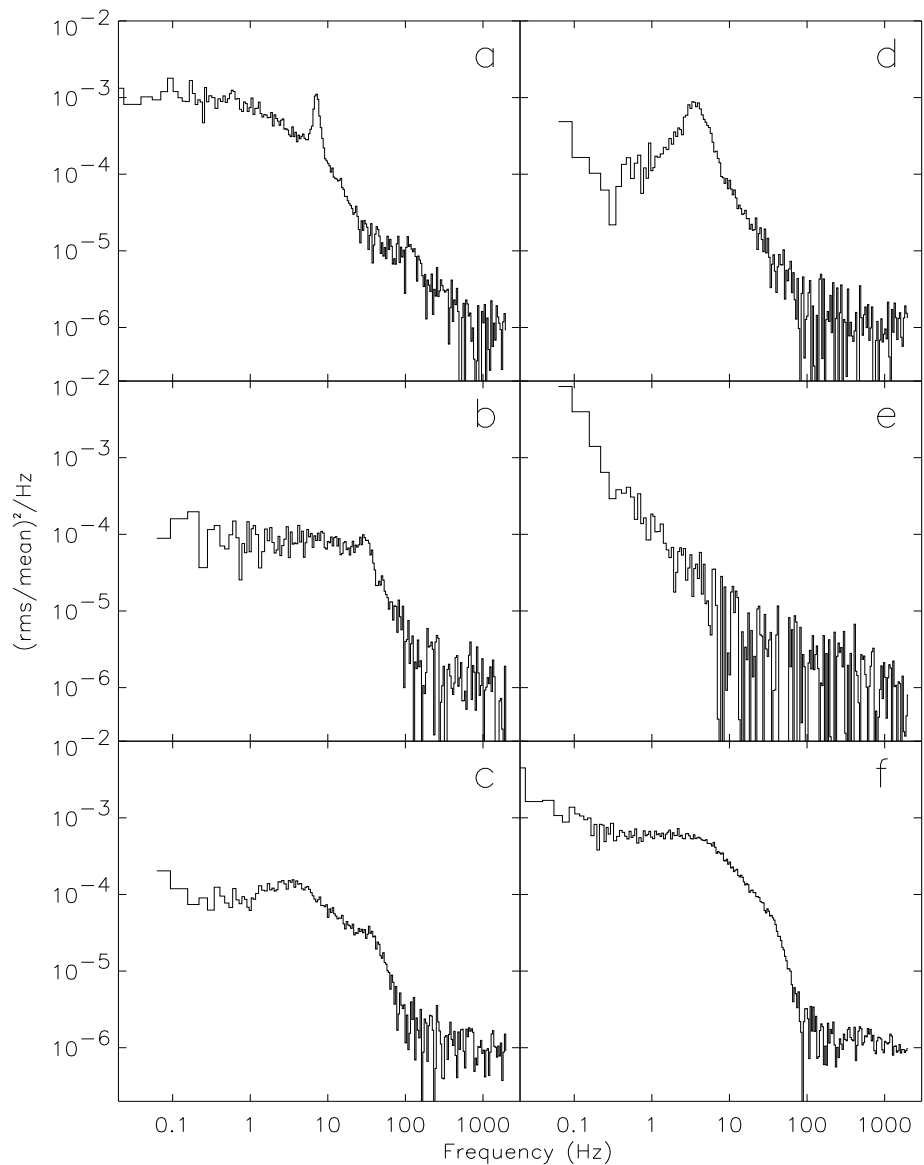,height=16.0cm,width=13.0cm,angle=+00}
\caption{
Power density spectra for six 1000-s intervals in
the 1997 Feb.-Mar cycle of Cir~X-1, shown in order of postulated
increase in accretion rate (i.e., more or less time reversed):
(a) the variable 1-12 Hz QPO seen at 7-Hz during quiesence
in Obs. VIII, (b) the same QPO in Obs. VI where it is found at 30-Hz, (c)
The constant 4-Hz QPO with a weakened 30 Hz `later' in Obs. VI.
(d) strong 4-Hz in Obs. V, (e) VLFN `later' in Obs. V, (f)
strong unpeaked noise in Obs. I. The latter may be a variant of (c); 
see note in text.}
\label{fig:4}
\end{figure*}

\subsection{Quasi-periodic oscillations}

The RXTE data from the eight 6-ksec observations during the long
descent of 1997 Feb. - Mar. clarify significantly the behavior of the
QPO's in Cir~X-1. As noted, QPO at 5--20~Hz were apparent in EXOSAT
data. Oosterbroek et~al. associated these with normal-branch
oscillations that are associated with instabilities in the accretion
flow under conditions of high luminosity (Fortner, Lamb \& Miller,
1989).

During the quiescent phase, Shirey et~al. reported a QPO with
frequency moving systematically from 1.3 to 12 Hz as well as other
features in the PDS. Additional RXTE data now show that this QPO moves
all the way up to 30 Hz (Fig.~4a,b). In addition, during the active
state, a strong 4-Hz oscillation can develop in concert with the
fading 30-Hz oscillations (Fig.~4c) and then can become quite strong
(Fig.~4d). The 4-Hz QPO can give way to strong very-low-frequency
noise due to active flaring and dips (Fig.~4e) or to strong unpeaked
noise (Fig.~4f), all during the active state. The 4-Hz oscillation
always appears at the same frequency and is often apparent. (Recent
analyses by Shirey suggest that Fig.~4f may be a variant of Fig.~4c
which occurs near the HB-NB apex; see below.)

The 1--30~Hz QPO frequency increases very strongly with 2--18~keV
intensity during the relatively small changes of intensity during
quiescence. In contrast, the 4-Hz QPO occurs over a wide range of
intensities and is quite constant. The latter QPO is reminiscent of
``normal branch'' oscillations in the LMXB sources exhibiting `Z'
behavior in the color-color (CCD) and hardness-intensity (HID)
diagrams (Hasinger \& van der Klis 1989). An example is the 6-Hz QPO
in Sco X-1 (Middleditch \& Priedhorsky 1986).  This further suggests
that the 1-30 Hz QPO are ``horizontal-branch'' QPO in the same scheme.
Indeed, the frequency increases with intensity as do horizontal branch
oscillations. Such oscillations are indicative of a significant
magnetic field in beat-frequency models (Alpar \& Shaham 1985; Lamb
et~al. 1985).

\subsection{Branches in the HID}

Color-color diagrams (CCD) and hardness-intensity diagrams (HID) are
shown in Fig.~5 for the same long descent from the active state in
1997 Feb. - Mar. The left two panels show all 8 observations, numbered
in time sequential order from 1 to 8, while the right two are blowups
of Observations~V and VI which occurred just as the source was making
the transition from the active to quiescent state at the end of the
long descending ramp seen in Fig.~1. Each point represents 16~s of
data. The scatter plot for each of the eight PCA observations tends to
be isolated from that of the adjacent observation because of the large
intensity and color changes.

\begin{figure*}
\epsfig{figure=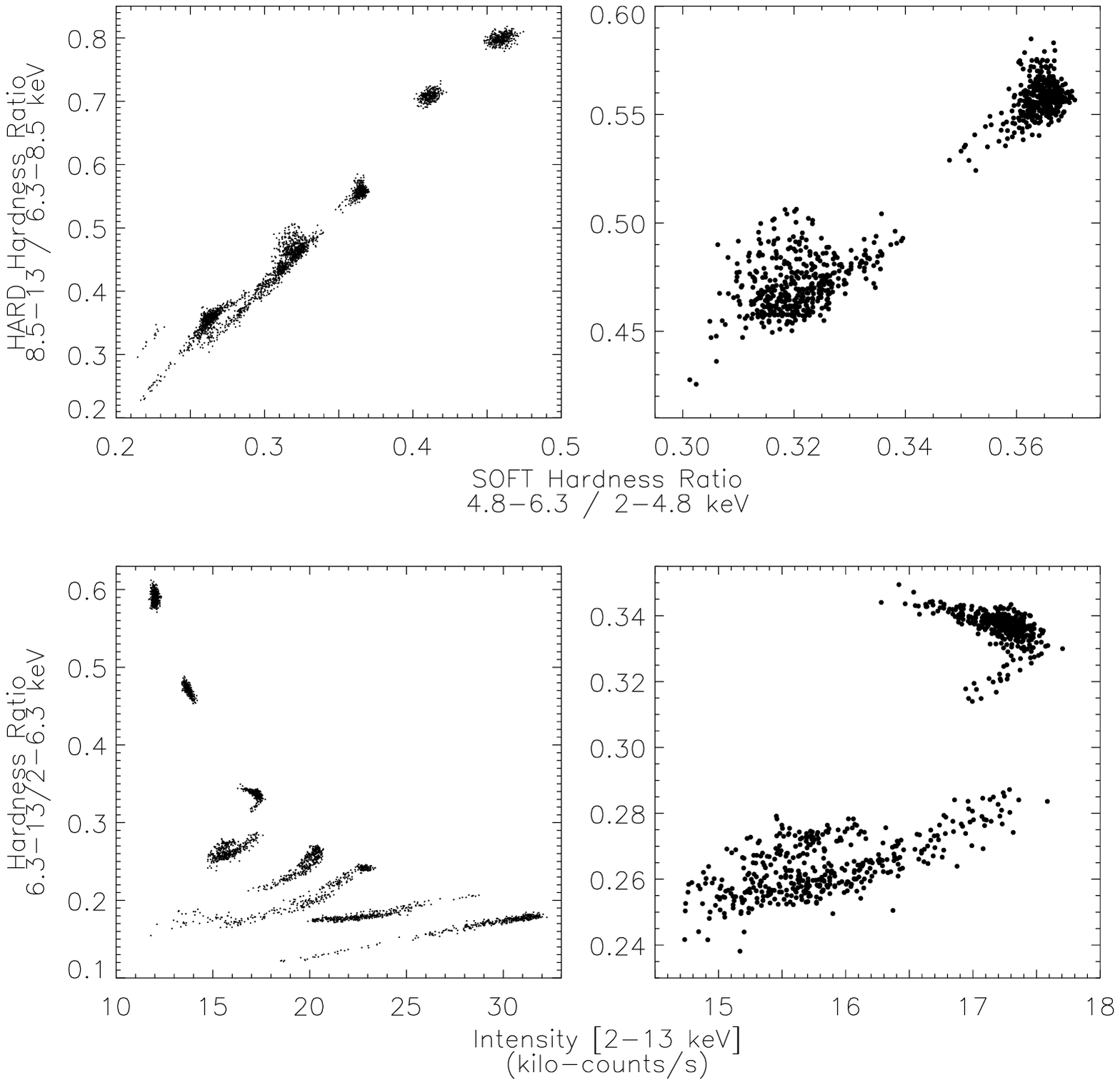,height=16.0cm,width=15.0cm,angle=+00}
\caption{
Left panels: all 8 observations of the 1997 Feb-Mar cycle
shown in a CCD (upper panel) and in an HID (lower panel). In time
order the observations (numbered I - VIII) progress from lower left to
upper right in the CCD and from lower right to upper left in the
HID. Right panels: expanded views of Obs. V and VI in CCD and HID. The
expanded HID (lower right panel) bears similarities to the appearance
of Z sources, particularly when the QPO characteristics are taken into
account.}
\label{fig:5}
\end{figure*}

The HID (lower panels) serve to isolate trends better than the CCD
(upper panels), so our discussion here will focus on them. One
immediately notes the long tails due to dipping and flaring in the
earlier observations (I - IV, lower right of lower-left panel), and
the lack of flaring in the last two observations (VII - VIII, upper
left of same panel).  The intermediate observations (V and VI) show
interesting branch structure; see lower right panel.

The shape of these two observations taken together is reminiscent of
Z sources. The data in this scatter plot were divided into 11 
separate zones and FFT were taken
for each of them. Several of the resultant PDS are shown in Fig.~4;
they will be referenced in the following discussion. 

Focus now on the lower right panel of Fig.~5 (HID for Obs. V and VI).
The variable QPO is seen at its highest frequency (30-Hz; Fig.~4b) at
the upper left end of the upper cluster (Obs. VI). Further to the
right along this ``horizontal branch'', the 30-Hz is fading as the 4-Hz
QPO begins to appear (Fig.~4c). The 4~Hz QPO remains strong around the
bend and onto the short descending "normal" branch without evidence of
the 30 Hz (Fig.~4d). In the lower cluster (Obs. V), the 4~Hz QPO is
present along the (longer) lower branch (continuation of the ``normal''
branch?).  Only large low frequency power is found on the upper
(``flaring''?)  branch of this observation (Fig.~4e).

This correlation of HID branches with QPO type reinforces the
association with Z sources. The postulated horizontal-branch
QPO does indeed occur on a horizontal branch in the HID, and
the postulated normal-branch QPO does occur on a descending branch as
in Z sources. The expected variation of the 1--30~Hz QPO is not seen
in the horizontal branch of Obs. VI, but this QPO appears at lower
frequency at lower accretion rates, e.g. 7~Hz in
Obs. VIII (Fig.~4a). 

Since this talk was given, Shirey has found that a similar diagram,
but with the hard color (13--18~keV/9--13~keV) on the ordinate
(instead of the soft color shown here), the upper tail of the lower
diagram turns downward rather than upward. Thus the perceived sequence
from top to bottom (say, as $\dot{M}$ increases) makes contiguous the
4-Hz regions, and places the ``flaring'' region with no QPO and strong
low-frequency noise below (`after') the 4-Hz branch.  This makes the
Z-source association even stronger. There do remain differences
between these diagrams and classic Z-source behavior, e.g., no Z shape
in the CCD). These differences may well be related to the extreme
degree of variability in this source.

\section{CONCLUSIONS}

The behavior of intensity, colors, and QPO during a long decrease in
intensity has shown behavior clearly reminiscent of low-mass Z
sources.  The presence of horizontal-branch oscillations in the
beat-frequency model (Alpar \& Shaham; Lamb et~al.)  indicates a
significant magnetic field. Clearly, the suggested low-magnetic-field
description of the neutron star in this system (Oosterbroek et~al.)
should be reexamined.

The timing and nature of absorption dips has been elucidated.  The
pronounced dips typically occur just as the source is entering its
active state, but before it has increased in intensity. These
particular dips have been shown in PCA data to be absorption events
with partial covering, confirming earlier work, but placing the
absorption clearly within the well-defined dips at the entry to the
active phase. These dips should be useful in unraveling the geometry
of this unique and interesting system.

The authors thank Drs.\ Edward Morgan and Saul Rappaport for helpful
conversations. They are also grateful to all those on the ASM and PCA
teams at MIT and GSFC, to the RXTE control and data management teams
at GSFC, and to NASA.

\section{REFERENCES}

Alpar, M. \& Shaham, J. 1985, Nature 316, 239

Brandt, W. et~al. 1996, MNRAS, 283, 1071

Fortner, B., Lamb, F., \& Miller, G. 1989, Nature 342, 775

Glass, I. 1978, MNRAS, 183, 335

Glass, I. 1994, MNRAS, 268, 742

Hasinger, G. \& van der Klis, M. 1989, A \& A 225, 79

Kaluzienski et~al. 1976, ApJ, 208, L71

Lamb, F. et~al. 1985, Nature 317, 681

Middleditch, J. \& Priedhorsky, W. 1986, ApJ 306, 230

Moneti, A. 1992, A\&A, 260, L7

Oosterbroek, T. et~al. 1995, A\&A, 297, 141

Predahl, P. \& Schmitt, J. 1995, A\&A 293, 889

Shirey, R., et~al. 1996, ApJ, 469, L21

Stewart, R., et~al. 1991, MNRAS, 253, 212

Stewart. R. et~al. 1993, MNRAS 261, 593

Tennant, A. Fabian, A., \& Shafer, R. 1986, MNRAS, 221, 27p

Tennant, A. 1987, MNRAS, 226, 971

Tennant, A. 1988, MNRAS, 230, 403

Tsunemi, H. 1989, PASJ, 41, 391

van der Klis, M. 1994, ApJS, 92, 511

Whelan, J. 1977, MNRAS, 181, 259.

\end{document}